\documentstyle[aps,epsf]{revtex}
\def\Journal#1#2#3#4{{#1} {\bf #2}, #3 (#4)}
\def\PRL{\em Phys. Rev. Lett.}
\def\PRD{{\em Phys. Rev.} D}
\def\Qplm#1{{Q^{\rm even}_{\,#1}}}
\def\Qclm#1{{Q^{\rm odd}_{\,#1}}}
\def\ddt{{\partial \over \partial t}}
\def\rt{{r_*}}
\begin{document}
\title{A slightly less grand challenge: Colliding Black
Holes using perturbation techniques}
\author{Hans-Peter Nollert$^{1,2}$,
        John Baker$^1$,
        Richard Price$^3$, and
        Jorge Pullin$^1$ \\
1. {\it Center for Gravitational Physics and Geometry, Department of
Physics\\The Pennsylvania State University, 104 Davey Lab, University
Park, PA 16802}\\
2. {\it Institut f\"ur Astronomie und Astrophysik, Universit\"at
T\"ubingen, Auf der Morgenstelle 10, 72076 T\"ubingen, Germany} \\
3. {\it Department of Physics, University of Utah, Salt Lake City, UT
84112}\\
}
\maketitle
\begin{abstract}
Perturbation techniques can be used as an
alternative to supercomputer calculations in calculating gravitational
radiation emitted by colliding black holes, provided the process
starts with the black holes close to each other. We give a summary of
the method and of the results obtained for various initial
configurations, both axisymmetric and without symmetry: Initially
static, boosted towards each other, counter-rotating, or boosted at an
angle (pseudo-inspiral). Where applicable, we compare the perturbation
results with supercomputer calculations.
\end{abstract}

\section{Introduction}
\begin{figure}[bth]
\hbox to \hsize{\epsfxsize=0.49\hsize\epsfbox{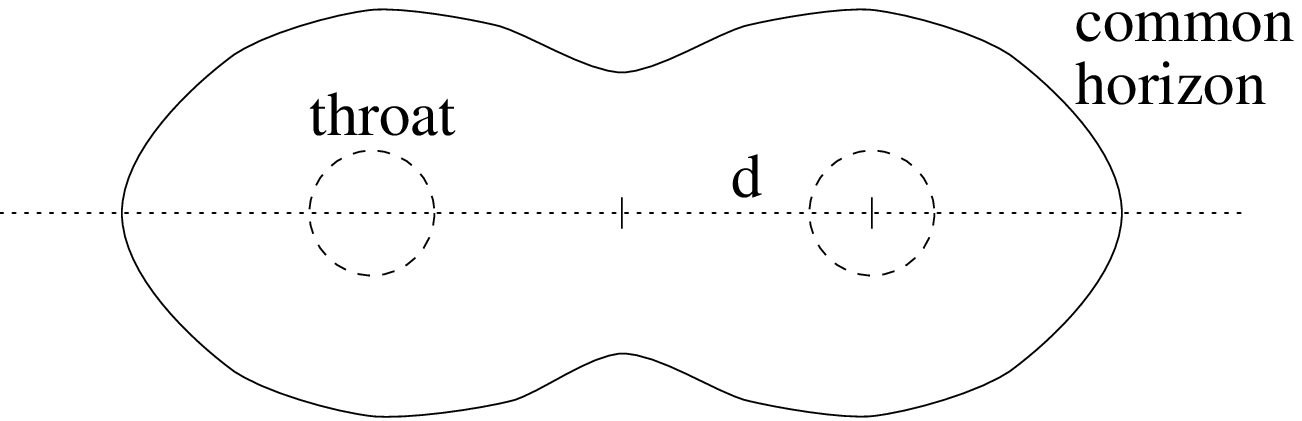}\hfil
      \epsfxsize=0.49\hsize\epsfbox{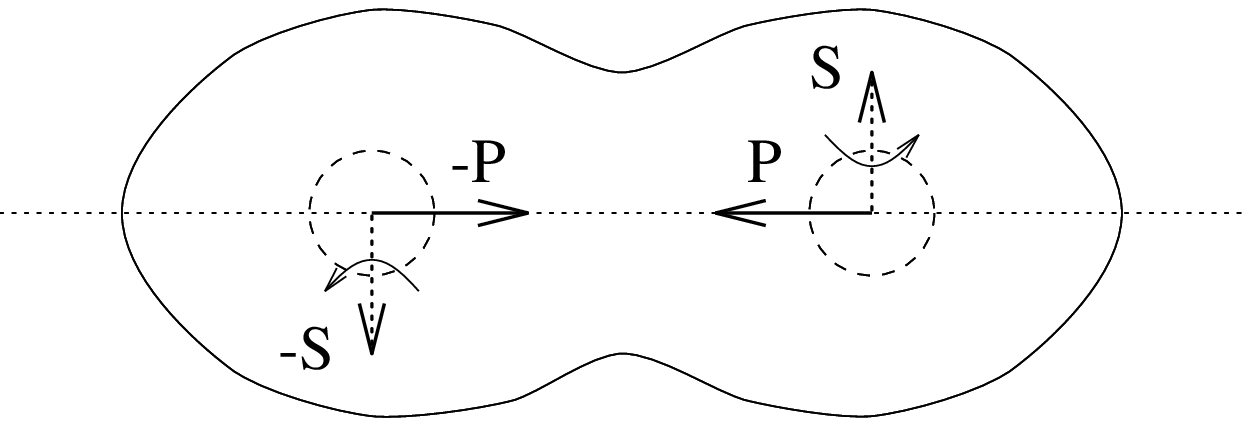}}
\smallskip\noindent
\centerline{\vbox{\hsize=0.82\hsize
\caption[Text to appear in the list of figures]
{\hsize=0.9\hsize\noindent
Regarding a close collision of two black holes as a 
perturbation of the final black hole. Left: Initially static. Right:
Initially boosted towards each other, including non-axisymmetric,
antiparallel spins.}}}
\label{1+1=1}
\end{figure}

The grand challenge for supercomputers in numerical relativity is the
simulation of the inspiral and merger of two black holes, and the
computation of the gravitational radiation emitted in the process. If
this event is to be described in full relativistic glory, the use of
supercomputers is inevitable. An alternative to this approach is the
use of perturbation theory: If the colliding black holes start out so
close to each other that they have already formed a common horizon,
they can be treated as a linearized perturbation of a single,
spherically symmetric black hole, as illustrated in
Fig.~\ref{1+1=1}. This method is less complicated and requires
considerably less computer resources (CPU time, memory), than the
full, non-linear approach. Furthermore, additional analytic insight
can be gained which is not accessible in the supercomputer
computation.

\section{The perturbation approach}

The full spacetime metric is written as a background metric plus a
small perturbation:
\begin{equation}\label{perturb}
g_{\mu\nu} = g_{\mu\nu}^0 + h_{\mu\nu}  \;.
\end{equation}
We will use the Schwarzschild metric as the background metric
$g_{\mu\nu}^0$. The field equations are then linearized around the
background metric, yielding linear equations for the perturbation.

\subsection{Time evolution}

The spherical symmetry of the background metric allows the separation
of the angular variables, and eventually the reduction
of the perturbation equations to a single, one-dimensional wave
equation with a potential $V(\rt)$ for a scalar function, the Zerilli
function $Q(\rt)$:
\begin{equation}\label{wave-equ}
\left[ {\partial^2 \over \partial t^2} - {\partial^2 \over \partial \rt^2}
        + V(\rt)\right] Q = 0 \;.
\end{equation}

There is one such Zerilli function for each value of $l$ and $m$,
which are the usual parameters in the spherical harmonics used for the
angular dependence, and for even and odd parity perturbations. The
potential $V(\rt)$ (and thus the wave equation) depends on $l$ and the
parity, but not on $m$.

\subsection{Initial data}

In order to obtain initial data for the black hole collision, we use
the approach described by Bowen and York \cite{BY80} solving the
constraint equations assuming a conformally flat initial spatial
slice:
\begin{equation}
g_{ij} = \phi^4 \eta_{ij}  \;.
\end{equation}
Solutions for the conformal factor $\phi$ describing several initially
static wormholes were found by Misner \cite{misner}. 

For configurations where the black holes are not static initially, we
use the solutions given by Bowen and York \cite{BY80} for the
conformal extrinsic curvature
\begin{eqnarray}\label{extcurv}
\widehat{K}_{ij} &=& {3 \over  2 r^2} \left[ 2 P_{(i} n_{j)} -
                       (\eta_{ij} - n_i n_j) P^c n_c \right] \\
\widehat K_{ij} &=& {3 \over r^3} \left(\epsilon_{kil} S^l n^k n_j +
                                        \epsilon_{kjl} S^l n^k n_i\right)
  \;,
\end{eqnarray}
valid for one black hole with momentum $P$ or spin $S$.
The extrinsic curvature for several black holes can be obtained by
combining contributions from each black hole. 

If the initial configuration is not static, then the conformal factor
will be different from its Misner form. However, the difference is of
second order in the extrinsic curvature. We can thus ignore the
difference, using the Misner form, as long as the extrinsic curvature
is not too large.

Initial data for the Zerilli function itself is derived from the
initial spatial metric \cite{PP}, the initial derivative is obtained
from the extrinsic curvature \cite{AP}.

\subsection{Radiation emitted}

Once the time evolution of all relevant Zerilli functions is known,
the total energy radiated as gravitational radiation during the
close collision is given by \cite{AP}
\begin{equation}\label{totalrad}
P(r\rightarrow\infty) = {1 \over 32 \pi} \sum_{l,m} \left(
                        \left(\ddt \Qplm{lm}\right)^2 +
                        \left(\Qclm{lm} \right)^2 \right)
\end{equation}

\section{Results}

\subsection{Axisymmetric collisions}

\begin{figure}[bth]
\begin{minipage}{0.40\hsize}
\epsfxsize=\hsize\epsfbox{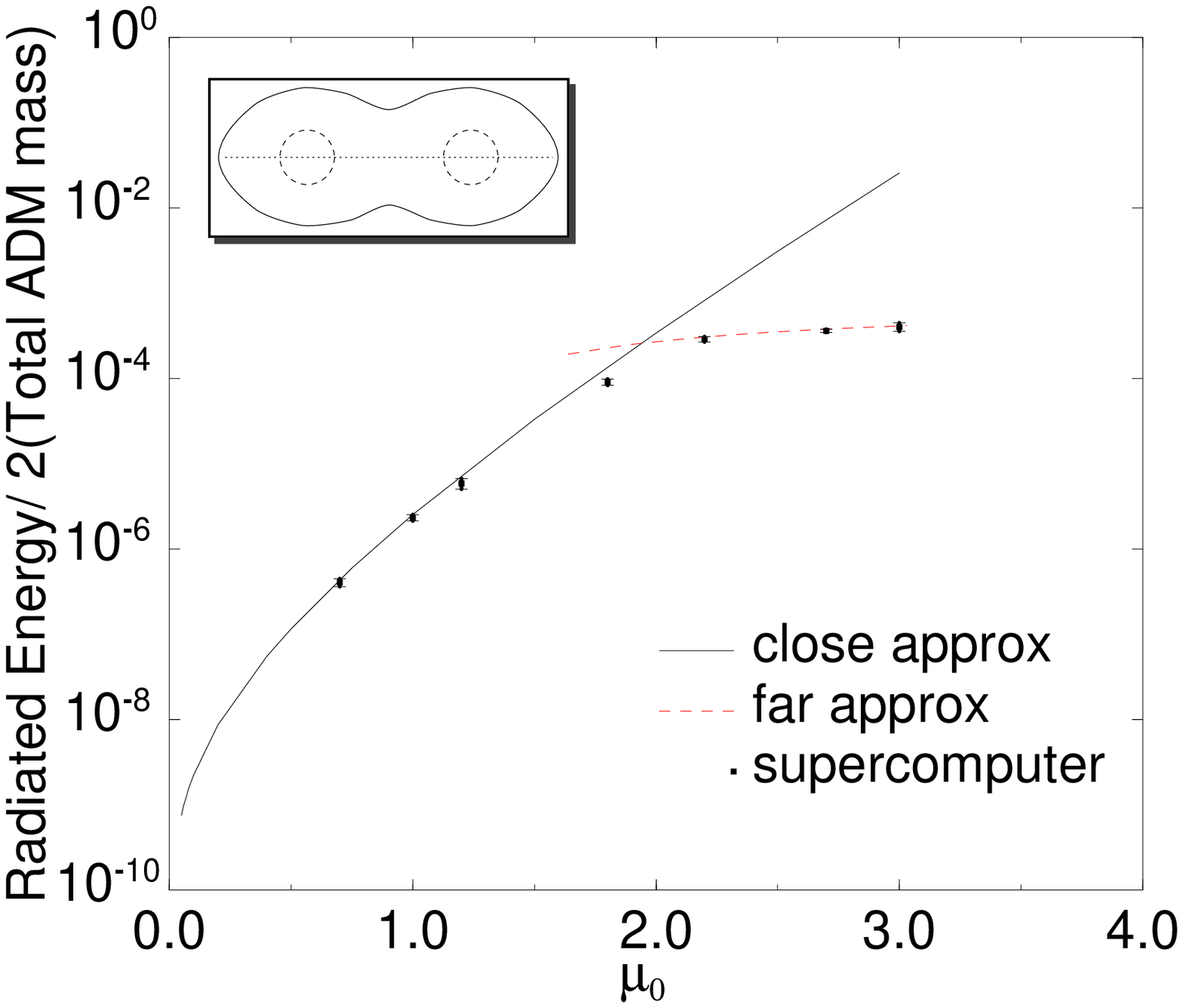}
\smallskip\noindent
\caption[Text to appear in the list of figures]
{\hsize=0.9\hsize\noindent
Total radiated energy, according to the close approximation, for
a collision of two initially static black holes. The parameter $\mu_0$
determines the initial separation.
}
\label{inistat}
\end{minipage}%
\hfil
\begin{minipage}{0.55\hsize}
\epsfxsize=\hsize\epsfbox{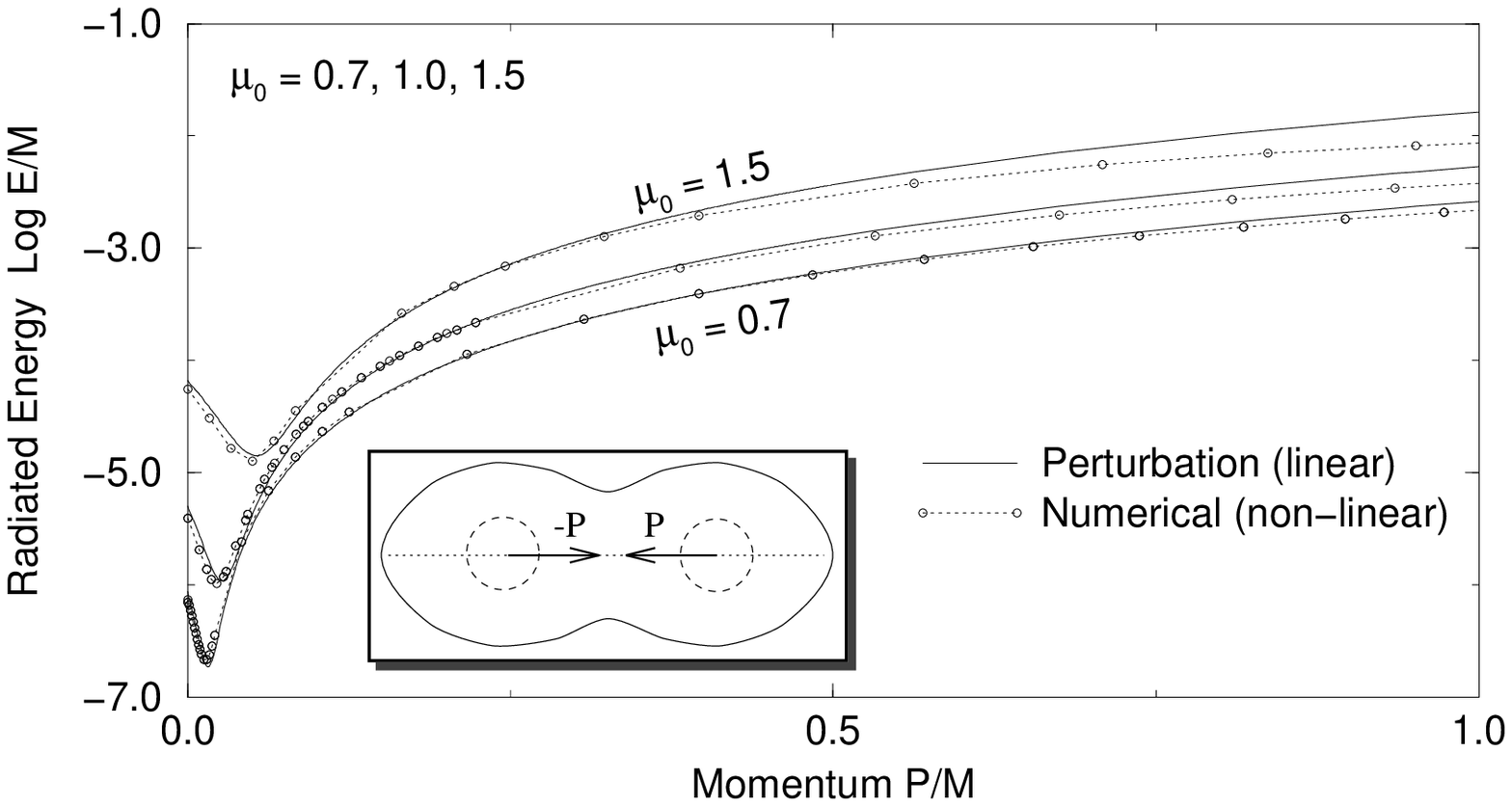}
\smallskip\noindent
\caption[Text to appear in the list of figures]
{\hsize=0.9\hsize\noindent
Total radiated energy for two black hole initially boosted towards
each other.}
\label{iniboost}
\end{minipage}
\end{figure}

The perturbation approach for close collisions was first used by
Pullin and Price \cite{PP} for two initially static black holes. The
results, shown in Fig.~\ref{inistat}, indicated excellent agreement
with supercomputer calculations for much larger separations than
expected. Even at $\mu_0$ = 1.36, where a common apparent horizon
only begins to form, the difference is less than a factor of 2.

Figure~\ref{iniboost} shows the radiated energy of two black holes
initially boosted towards each other \cite{boost}. It turns out that
the agreement with fully non-linear, numerical calculations also
extends to rather large values of the initial momentum. This is again
rather unexpected, since the construction of initial data assumed
small extrinsic curvature, and thus small initial momentum.

Two counterrotating black holes, with antiparallel spins aligned along
the direction of the collision (``Cosmic Screw''), still form an
axisymmetric system and result in a final Schwarzschild black
hole. The radiated energy as a function of spin is shown in
Fig.~\ref{screw}. If both black holes have a spin of $S = \pm 1 M^2$,
the rdaiated energy is incerased by about a factor of $400 \dots 1000$
over the non-rotating case.

\subsection{Non-axisymmetric collisions}

Other than for supercomputer calculations, axisymmetry does not play a
major role in the perturbation approach. Two black holes with
antiparallel spins, aligned perpendicular to the line of collision,
form a simple non-axisymmetric system. It is usually expected that the
violation of axisymmetry will increase the amount of radiation. This
is not true for this particular situation; rather, the radiation
contributed by the spins is {\it lower} by a factor of $3/4$, compared
to the axisymmetric case.

\begin{figure}[bth]
\begin{minipage}[t]{0.475\hsize}
\epsfxsize=\hsize\epsfbox{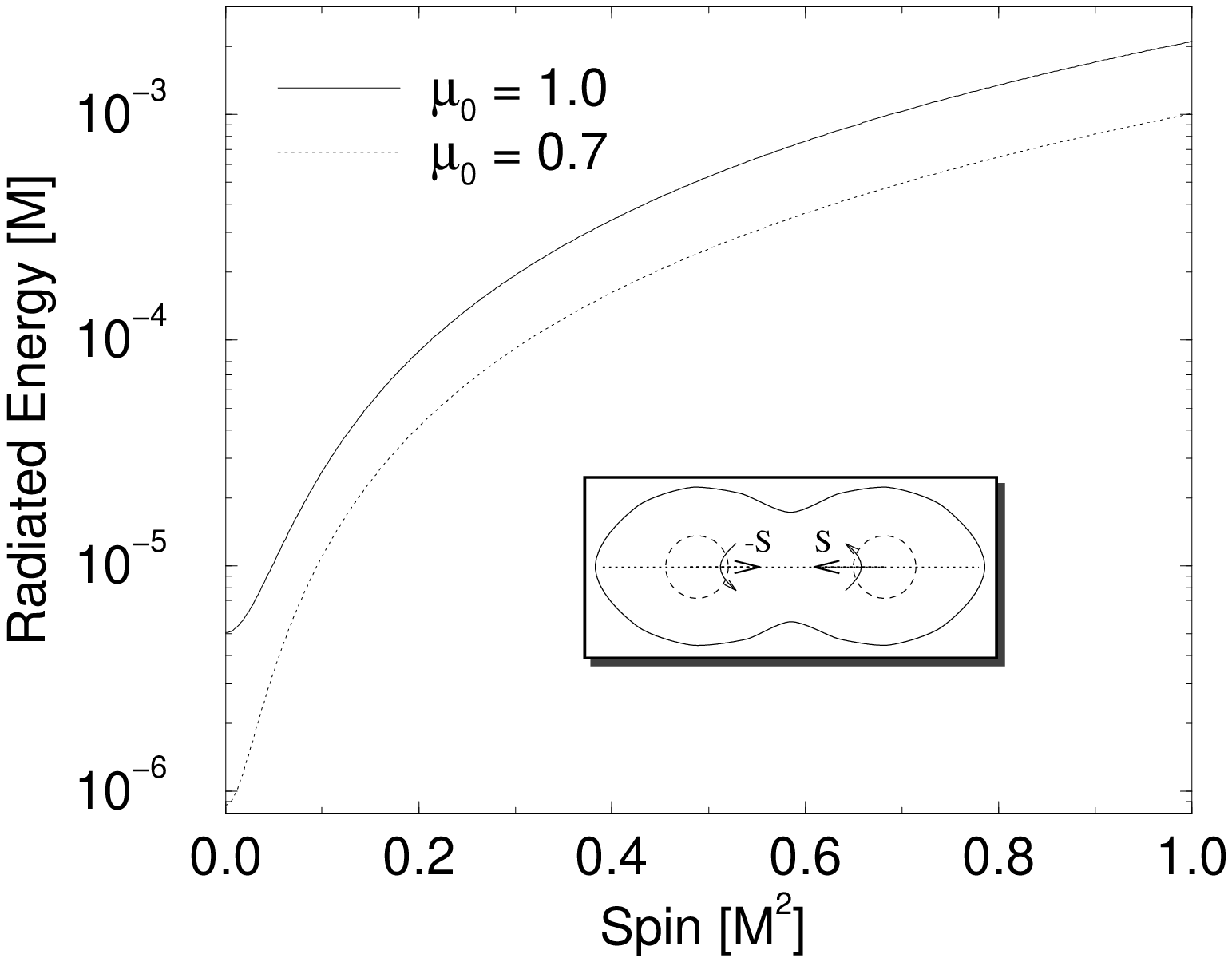}
\smallskip\noindent
\caption[Text to appear in the list of figures]
{\hsize=0.9\hsize\noindent
Total radiated energy for two black holes carrying antiparallel spins 
aligned with the direction of collision (``Cosmic Screw'').}
\label{screw}
\end{minipage}%
\hfil
\begin{minipage}[t]{0.475\hsize}
\epsfxsize=\hsize\epsfbox{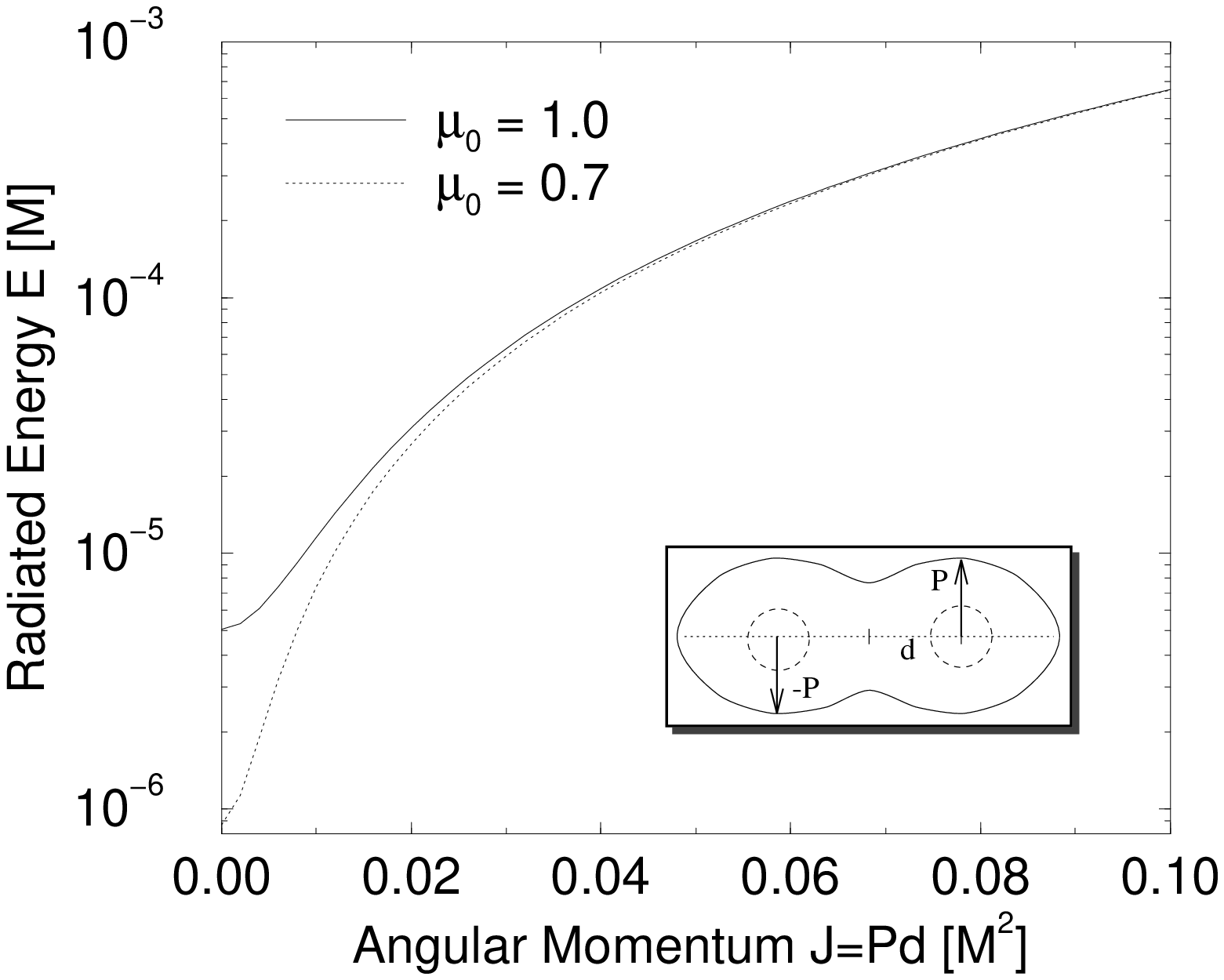}
\smallskip\noindent
\caption[Text to appear in the list of figures]
{\hsize=0.9\hsize\noindent
Total radiated energy for black hole collisions with the
black holes being initially boosted perpendicularly to the line
connecting them (``Pseudo-Inspiral'').}
\label{insp}
\end{minipage}%
\end{figure}

Boosting the black holes perpendicularly to the line connecting them
also violates axisymmetry. This configuration can be regarded as the
start of a collision from a slowly decaying circular orbit. 
Figure~\ref{insp} shows the radiated energy as a function of the
resulting orbital angular momentum. Since this case results in a
rotating black hole, we can only treat it as a perturbation of a
Schwarzschild black hole as long as the final angular momentum is
small. However, even for $J=0.1M^2$, the emitted radiation increases
by about two orders of magnitude, compared to the head-on collision.

\section*{Acknowledgments} This work was supported in part by grants
NSF-PHY-9423950, NSF-PHY-9514240, NSF-PHY-95-07719, NSF-PHY-91-16682,
NSF-PHY-9404788, NSF-ASC/PHY-93-18152 (ARPA supplemented), and by the
DFG through grant No~219/5-2.


\begin{thebibliography}{99} 
\bibitem{BY80}
J.M. Bowen and J.W. York, Jr., \Journal{\PRD}{21}{2047}{1980}
\bibitem{misner}
C.W. Misner, \Journal{Ann. Phys.}{24}{102}{1963}
\bibitem{PP}
R.H.  Price and J. Pullin, \Journal{\PRL}{72}{3297}{1994}
\bibitem{AP}
A.M. Abrahams and R.H. Price, \Journal{\PRD}{53}{1963}{1996}.
\bibitem{boost}
J. Baker {\it et al}, \Journal{\PRD}{55}{829}{1997}.
\end{thebibliography}
\end{document}